\documentclass[conference]{IEEEtran}
\usepackage{cite,graphicx,hyperref}
\usepackage{amsfonts,amsmath,amssymb}
\usepackage{xcolor}
\usepackage{bm}
\usepackage{algorithm,algpseudocode}
\hypersetup{colorlinks=true, citecolor=blue, linkcolor=blue, urlcolor=cyan}
\definecolor{azure}{rgb}{0.0, 0.5, 1.0}
\definecolor{frenchblue}{rgb}{0.0, 0.45, 0.73}
\definecolor{forestgreen(traditional)}{rgb}{0.0, 0.27, 0.13}
\definecolor{mygreen}{rgb}{0.09, 0.45, 0.27}
\definecolor{myblue}{rgb}{0.2383,0.5195,0.7734}
\definecolor{mygreen}{rgb}{0.6445,0.9297,0.0039}
\definecolor{darklavender}{rgb}{0.45, 0.31, 0.59}
\definecolor{americanrose}{rgb}{1.0, 0.01, 0.24}
\definecolor{pigblue}{rgb}{0.2, 0.2, 0.6}
\definecolor{blue(ryb)}{rgb}{0.01, 0.28, 1.0}
\definecolor{amethyst}{rgb}{0.6, 0.4, 0.8}
\definecolor{deepmagenta}{rgb}{0.8, 0.0, 0.8}
\definecolor{carminered}{rgb}{1.0, 0.0, 0.22}
\definecolor{iris}{rgb}{0.35, 0.31, 0.81}
\newcommand*{\herm}{{\mkern-1.5mu\mathsf{H}}}

\def\ccal0{{\ensuremath{\mathcal 0}}}
\def\bb0{{\ensuremath{\boldsymbol 0}}}

\title{Covariance-Aware MM-PGD for Mixed Near-/Far-Field Activity Detection}

\author{
\IEEEauthorblockN{Xinjue~Wang\IEEEauthorrefmark{1}, Zhi-Yong~Wang\IEEEauthorrefmark{2}, Sergiy~A.~Vorobyov\IEEEauthorrefmark{1}, Esa~Ollila\IEEEauthorrefmark{1}, \\
Gayan~Amarasuriya~Aruma~Baduge\IEEEauthorrefmark{3}, and Mojtaba~Vaezi\IEEEauthorrefmark{4}
\thanks{This work was supported by the Research Council of Finland under Grant 359848.}}
\IEEEauthorblockA{\IEEEauthorrefmark{1}Department of Information and Communications Engineering, Aalto University, 02150 Espoo, Finland\\}
\IEEEauthorblockA{\IEEEauthorrefmark{2}State Key Laboratory of Ocean Sensing, Ocean College, Zhejiang University, Hangzhou 310000, China\\}
\IEEEauthorblockA{\IEEEauthorrefmark{3}School of ECBE, Southern Illinois University, Carbondale, IL 62918 USA}
\IEEEauthorblockA{\IEEEauthorrefmark{4}ECE Department, Villanova University, Villanova, PA 19085 USA\\
}

E-mail: xjw@ieee.org, zhiyong.wang@zju.edu.cn, \{sergiy.vorobyov, esa.ollila\}@aalto.fi, \\gayan.baduge@siu.edu, mvaezi@villanova.edu
}

\newcommand\blfootnote[1]{%
  \begingroup
  \renewcommand\thefootnote{}\footnote{#1}%
  \addtocounter{footnote}{-1}%
  \endgroup
}

\begin{document}
\maketitle
\blfootnote{This work was supported by the Research Council of Finland under Grant 359848. The work of Gayan Aruma Baduge in part was supported by the National Science Foundation (NSF) under Grant CCF-2326621.
The work of Mojtaba Vaezi was supported by the NSF under Grant CCF-2326622.}

\begin{abstract}
Grant-free activity detection  with mixed near-field (NF) and far-field (FF) devices is an important problem that can be addressed via covariance-based detectors. 
The difficulty is that NF users induce device-specific structured spatial covariances, whereas FF users are well approximated by isotropic covariances. 
Under a unified Rician model, we first formulate activity detection as a relaxed maximum-likelihood problem in the full $LM$-dimensional vectorized observation space. 
We then develop a covariance-aware majorization--minimization projected gradient descent (MM-PGD) detector. 
It updates the full activity vector jointly and avoids the per-coordinate high-rank subproblems that arise in the NF regime. 
Numerical performance results versus SNR, antenna-count, and NF-ratio sweeps show that MM-PGD achieves up to $20\times$ lower miss-detection probability than the strongest coordinate-wise baseline. 
The advantage is most pronounced at high NF ratios, while in the all-FF case MM-PGD performs on par with the strongest coordinate-wise baseline.
\end{abstract}

\section{Introduction}
\label{sec:intro}

Grant-free random access is a central mechanism for massive machine-type communications, where only a small subset of registered devices is active in each coherence block and the base station must infer the active set from short uplink pilots~\cite{shahab2020grant,chen2020massive,liu2018sparse,bockelmann2018towards,senel2018grant,chen2018sparse}.
Classical activity-detection methods such as approximate message passing (AMP)~\cite{liu2018massiveAMP}, sparse Bayesian learning (SBL)~\cite{wipf2004sparse}, and covariance-based coordinate-wise optimization~\cite{fengler2021nonTIT} mainly target far-field (FF) uncorrelated fading case, where the antenna dimension can be treated as repeated snapshots.

With large arrays, however, a non-negligible part of the device population may operate in the radiating near field (NF)~\cite{Liu2023NFReview}.
Then the channel covariance is no longer isotropic. Spherical wavefronts generate device-dependent structured spatial covariance, while more distant devices remain effectively far-field.
This mixed NF/FF regime breaks the standard snapshot approximation and calls for detectors that operate directly on the joint space--time covariance of the vectorized observation~\cite{Liu2023NFReview,wangcas2025TWCNearField}.
We adopt the standard statistical-CSI viewpoint of covariance-based activity detection (AD). 
The long-term channel mean and covariance of the registered device pool are assumed available at the base station (BS) and learned offline from historical uplink observations or geometry-based models.

\subsection{Related Work}

Most covariance-based grant-free detectors are built around the FF snapshot model, which enables coordinate-wise covariance learning and sparse-recovery updates with relatively simple subproblems~\cite{fengler2021nonTIT,marata2024activity,chen2021phase}.
The same FF modeling regime also underlies MMV-type AMP and SBL formulations~\cite{liu2018massiveAMP,wipf2004sparse}.
The covariance-based viewpoint has also been extended beyond the basic single-cell setting, for example to joint activity/data detection, cooperative multi-cell processing, and cell-free architectures~\cite{chen2019covarianceICC,wang2024covarianceMultiCellTIT,ganesan2021clustering}.
Robust covariance-based variants have also been studied to improve resilience under model mismatch and non-ideal observation conditions~\cite{wang2025RobustADTSP,wang2025ADicassp}.
Rician maximum-likelihood estimation (MLE) formulations with explicit line-of-sight (LoS) means have also been studied under isotropic covariance assumptions~\cite{Liu2024CWOMMLE}. 
They remain informative in the FF regime but do not retain the full NF covariance structure.
Recent NF work has extended this line to structured low-rank spatial covariances~\cite{wangcas2025TWCNearField}. 
The resulting solvers are still coordinate-wise, where per-user updates become substantially heavier because NF covariance destroys the FF rank-one update structure.

% Motivation, main differences between counterparts and chanllenge.
This paper instead considers a mixed NF/FF setup, where FF and NF devices coexist within the same AD model. 
The method in~\cite{Liu2024CWOMMLE} retains the LoS mean but collapses each device covariance to an isotropic scaled-identity form, whereas~\cite{wangcas2025TWCNearField} retains structured NF covariance through coordinate-wise high-rank subproblems. 
Here we instead formulate mixed NF/FF AD through a single covariance-aware likelihood and solve it by global majorization--minimization (MM) updates. 
The main difficulty is that exact likelihood evaluation remains coupled in the full $LM$-dimensional domain once the classical FF rank-one simplification disappears.

% ---
\subsection{Contributions}

In this paper, we focus on a compact single-cell detector for mixed NF/FF activity detection under representative operating points. 
The emphasis is on detector formulation, global MM-based updates, and detection performance rather than on large-scale implementation details.

The contributions of this paper are as follows.
First, we formulate the mixed NF/FF activity detection problem as relaxed maximum-likelihood estimation under a unified Rician model with device-dependent mean $\bar{\bm h}_n$ and covariance~$\bm R_n$.
Second, we propose a MM projected gradient descent detector that updates the full activity vector jointly and avoids the per-coordinate high-rank subproblems faced by coordinate-wise solvers in the NF regime, while covering FF, NF, and mixed NF/FF settings within one unified covariance-aware formulation. 
Finally, numerical results over SNR, antenna count, and NF-ratio sweeps, where the NF ratio denotes the fraction of NF devices in the registered device pool, show clear gains over four representative baselines, with the largest improvements appearing as the NF fraction grows.

\section{System Model}
\label{sec:model}

We consider an uplink single-cell multiple-input multiple-output (MIMO) system with a BS equipped with a uniform linear array of $M$ antennas and a registered pool of $N$ single-antenna devices, among which $K\ll N$ are active.
Each device $n$ is assigned a length-$L$ pilot $\bm s_n\in\mathbb{C}^{L}$ with $\|\bm s_n\|_2=1$.
The received signal matrix is modeled as
\begin{equation}
\bm Y=\sum_{n=1}^{N}\alpha_n \bm s_n \bm h_n^\top+\bm W,
\label{eq:Y}
\end{equation}
where $\alpha_n\in\{0,1\}$ is the activity indicator, $\bm h_n\in\mathbb{C}^{M}$ is the channel, $(\cdot)^\top$ denotes transpose, and~$\bm W$ contains i.i.d.\ $\mathcal{CN}(0,\sigma_w^2)$ entries~\cite{Liu2023NFReview}.

% Mixed NF/FF Rician Channel Model

We adopt the unified mixed NF/FF statistical model
\begin{equation}
\bm h_n \sim \mathcal{CN}(\bar{\bm h}_n,\bm R_n),
\label{eq:rician_unified}
\end{equation}
where the pair $(\bar{\bm h}_n,\bm R_n)$ depends on whether device $n$ is in the NF or FF of the array~\cite{Liu2023NFReview,wangcas2025TWCNearField}.

For an NF device, we use a spherical-wave multipath model
\begin{equation}
\bm h_n=\beta_{n,0}\bm b(r_n,\theta_n)+\sum_{\ell=1}^{L_n}\beta_{n,\ell}\bm b(r_{n,\ell},\theta_{n,\ell}),
\label{eq:physical_channel}
\end{equation}
where $\beta_{n,0}$ is the LoS gain, and $\beta_{n,\ell}$, $r_{n,\ell}$, and $\theta_{n,\ell}$ denote the gain, distance, and angle of the $\ell$-th scatterer.
The near-field steering vector $\bm b(r,\theta)\in\mathbb C^M$ has entries
\begin{equation}
[\bm b(r,\theta)]_m=\frac{1}{\sqrt M}
\exp\!\left(-j\frac{2\pi}{\lambda}\big(d_m(r,\theta)-r\big)\right),
\label{eq:steering_vec}
\end{equation}
with
\begin{equation}
\begin{split}
d_m(r,\theta)&=\sqrt{r^2+\delta_m^2-2r\delta_m\sin\theta},\\
\delta_m&=\left(m-\frac{M+1}{2}\right)d,
\end{split}
\label{eq:dist_exact}
\end{equation}
where $d$ denotes the inter-element spacing of the uniform linear array. If the non-line-of-sight (NLoS) gains satisfy $\beta_{n,\ell}\sim\mathcal{CN}(0,\sigma_{n,\ell}^2)$ independently, then
\begin{equation}
\begin{split}
\bar{\bm h}_n&=\beta_{n,0}\bm b(r_n,\theta_n),\\
\bm R_n&=\sum_{\ell=1}^{L_n}\sigma_{n,\ell}^2\bm b(r_{n,\ell},\theta_{n,\ell})\bm b(r_{n,\ell},\theta_{n,\ell})^\herm.
\end{split}
\label{eq:cov_def}
\end{equation}
Hence $\bm R_n$ is a sum of $L_n$ rank-one terms and admits rank $r_n\triangleq \mathrm{rank}(\bm R_n)\le L_n<M$.

For FF devices, we use the standard isotropic specialization
\begin{equation}
\bm h_n\sim\mathcal{CN}(\bar{\bm h}_n,g_n\bm I_M),
\label{eq:ff_stats}
\end{equation}
where $g_n>0$ captures the large-scale fading.

We assume that the BS knows whether each registered device is NF or FF, e.g., from coarse geometry, range information, or sensing-assisted registration.
The detector therefore estimates only the activity variables $\{\alpha_n\}$.
We also assume that the long-term descriptors $\{(\bar{\bm h}_n,\bm R_n)\}_{n=1}^N$ are learned a priori, e.g., from covariance tracking over repeated uplink observations, geometry-based channel modeling, or low-rank NF covariance fitting based on estimated scatterer geometry~\cite{Liu2023NFReview,wangcas2025TWCNearField}.
This offline learning stage is more demanding for NF devices than for the FF isotropic model: the FF specialization only fits a scalar large-scale fading level, whereas the NF case must recover a structured covariance matrix, whose complexity grows with the array size $M$ and the effective scatterer count $L_n$.
Thus, the proposed detector is not a joint activity-and-environment learning scheme, and it does not re-estimate a fresh covariance matrix from extra pilots in every access slot.

\section{Problem Formulation}
\label{sec:problem}

\subsection{Vectorized Observation and Joint Statistics}

In classical all-FF models with $\bm R_n=g_n\bm I_M$, the $M$ antenna-domain observations can be treated as repeated snapshots.
This simplification breaks in mixed NF/FF settings because the NF covariances are generally non-isotropic and couple the antenna domain.
We therefore work directly with the vectorized observation
\begin{equation}
\bm y\triangleq \mathrm{vec}(\bm Y)\in\mathbb C^{LM}.
\end{equation}
Using $\mathrm{vec}(\bm u\bm v^\top)=\bm v\otimes \bm u$, we obtain
\begin{equation}
\bm y=\sum_{n=1}^{N}\alpha_n(\bm h_n\otimes \bm s_n)+\bm w,
\label{eq:vec_y}
\end{equation}
where $\bm w=\mathrm{vec}(\bm W)\sim\mathcal{CN}(\bm 0,\sigma_w^2\bm I_{LM})$.
Conditioned on $\bm\alpha$, $\bm y$ follows the Gaussian model
\begin{equation}
\bm y\mid \bm\alpha \sim \mathcal{CN}(\bm \mu_{\bm\alpha},\bm \Sigma_{\bm\alpha}),
\end{equation}
with mean and covariance
\begin{equation}
\bm \mu_{\bm\alpha}=\sum_{n=1}^{N}\alpha_n \bm m_n,\qquad
\bm \Sigma_{\bm\alpha}=\sum_{n=1}^{N}\alpha_n \bm C_n+\sigma_w^2\bm I_{LM},
\label{eq:stats}
\end{equation}
where $\bm m_n\triangleq \bar{\bm h}_n\otimes \bm s_n$ and $\bm C_n\triangleq \bm R_n\otimes(\bm s_n\bm s_n^\herm)$.

The per-device covariance matrix $\bm C_n$ inherits the spatial structure of $\bm R_n$:
\begin{itemize}
\item \emph{Far-field devices} ($\bm R_n=g_n\bm I_M$): $\bm C_n=g_n(\bm I_M\otimes \bm s_n\bm s_n^\herm)$ retains a Kronecker structure that decomposes into $M$ independent rank-one contributions in the $L\times L$ snapshot covariance domain.
\item \emph{Near-field devices}: by~\eqref{eq:cov_def}, $\bm R_n$ is a low-rank sum of steering-vector outer products, or equivalently $\bm R_n=\bm U_n\bm\Lambda_n\bm U_n^\herm$ with rank $r_n\le L_n<M$. Then $\bm C_n=\bm R_n\otimes(\bm s_n\bm s_n^\herm)$ no longer decomposes into independent snapshot-domain terms and instead couples the antenna and pilot dimensions.
\end{itemize}

In a mixed NF/FF population, the aggregate covariance $\bm\Sigma_{\bm\alpha}$ is therefore structurally heterogeneous: FF devices contribute Kronecker-structured terms while NF devices introduce cross-dimensional coupling.
%
% Offline/online split.
Given the long-term descriptors $(\bar{\bm h}_n,\bm R_n)$ and pilots $\bm s_n$, the per-device quantities $\bm m_n$ and $\bm C_n$ in~\eqref{eq:stats} are fixed and can be precomputed once for the registered user pool.
The online detector then uses only the current observation $\bm y$ together with the precomputed per-device mean vectors $\bm m_n$ and covariance matrices $\bm C_n$.

\subsection{Relaxed Maximum-Likelihood Detection}

Given $(\bm\mu_{\bm\alpha},\bm\Sigma_{\bm\alpha})$, the likelihood under a binary activity pattern $\bm\alpha$ is
\begin{equation}
p(\bm y\mid \bm\alpha)=
\frac{1}{\pi^{LM}|\bm\Sigma_{\bm\alpha}|}
\exp\!\left(
-(\bm y-\bm\mu_{\bm\alpha})^\herm \bm\Sigma_{\bm\alpha}^{-1}(\bm y-\bm\mu_{\bm\alpha})
\right).
\label{eq:likelihood}
\end{equation}
Dropping the constant $LM\log\pi$ yields the negative log-likelihood (NLL)
\begin{equation}
\mathcal{L}(\bm\alpha)\triangleq
\log|\bm \Sigma_{\bm\alpha}|
+
(\bm y-\bm \mu_{\bm\alpha})^\herm
\bm \Sigma_{\bm\alpha}^{-1}
(\bm y-\bm \mu_{\bm\alpha}).
\label{eq:nll_alpha}
\end{equation}
The exact ML detector solves a combinatorial problem over $\{0,1\}^N$.
To obtain a tractable formulation, we relax $\bm\alpha$ to $\bm\gamma\in[0,1]^N$ and reuse the same functional forms for $\bm\mu_{\bm\gamma}$ and $\bm\Sigma_{\bm\gamma}$.
This gives the relaxed ML problem
\begin{equation}
\min_{0\le \bm\gamma\le 1}\;
\mathcal{L}(\bm\gamma)\triangleq
\log|\bm \Sigma_{\bm\gamma}|
+
(\bm y-\bm \mu_{\bm\gamma})^\herm
\bm \Sigma_{\bm\gamma}^{-1}
(\bm y-\bm \mu_{\bm\gamma}).
\label{eq:problem}
\end{equation}

\section{MM-PGD Detector}
\label{sec:mmpgd}

\subsection{Why Full-Vector Updates}

Classical covariance-based detectors often solve~\eqref{eq:problem} by updating one coordinate $\gamma_n$ at a time.
Each such step perturbs the aggregate covariance as
\begin{equation}
\bm\Sigma_{\text{new}} = \bm\Sigma_{\text{old}} + \Delta\gamma_n\,\bm C_n.
\label{eq:coord_perturb}
\end{equation}
Under far-field isotropic fading, the Kronecker structure of $\bm C_n=g_n(\bm I_M\!\otimes\!\bm s_n\bm s_n^\herm)$ allows this update to decompose into $M$ independent rank-one inverse updates in the $L\!\times\! L$ snapshot covariance domain, which is the basis of efficient coordinate-wise solvers~\cite{fengler2021nonTIT}.
In the NF regime, however, $\bm C_n$ has rank $r_n>1$ and loses this FF simplification.
Each coordinate step must then act on the full $LM\times LM$ covariance, which becomes increasingly unattractive as the NF fraction grows.
This motivates a full-vector solver that updates $\bm\gamma$ jointly.

\subsection{Gradient of the NLL}
\label{subsec:gradient}

Define the residual $\bm r_{\bm\gamma}\triangleq \bm y-\bm\mu_{\bm\gamma}$ and the whitened residual $\bm v\triangleq \bm\Sigma_{\bm\gamma}^{-1}\bm r_{\bm\gamma}$.
Direct differentiation of $\mathcal{L}(\bm\gamma)=\log|\bm\Sigma_{\bm\gamma}|+\bm r_{\bm\gamma}^\herm\bm\Sigma_{\bm\gamma}^{-1}\bm r_{\bm\gamma}$ yields
\begin{equation}
[\nabla \mathcal{L}(\bm\gamma)]_n=
\underbrace{\mathrm{tr}(\bm \Sigma_{\bm\gamma}^{-1}\bm C_n)}_{\text{cov.\ fitting}}
-\underbrace{\bm v^\herm \bm C_n \bm v}_{\text{data corr.}}
-\underbrace{2\,\mathrm{Re}\{\bm v^\herm \bm m_n\}}_{\text{mean shift}}.
\label{eq:grad}
\end{equation}
The three terms admit intuitive interpretations: the first measures the expected contribution of device~$n$ to the covariance, the second measures the observed contribution via the whitened data, and the third accounts for the LoS mean.
Crucially, every component~$[\nabla\mathcal{L}]_n$ depends on $\bm v=\bm\Sigma_{\bm\gamma}^{-1}\bm r_{\bm\gamma}$, which couples all activity variables through~$\bm\Sigma_{\bm\gamma}$.

\subsection{Majorization--Minimization Update}
\label{subsec:mm_update}

\begin{algorithm}[t]
\caption{MM-PGD for Grant-Free Activity Detection}
\label{alg:mmpgd}
\begin{algorithmic}[1]
\Require Observation $\bm y$, pilots $\{\bm s_n\}$, channel statistics $\{(\bar{\bm h}_n,\bm R_n)\}$, and, for the oracle top-$K$ benchmarking protocol used here, active devices count $K$
\Ensure Detected active set $\hat{\mathcal{S}}$
\State Initialize $\bm\gamma^{(0)}\gets (K/N)\mathbf{1}_N$; set $t\gets 0$
\Repeat
    \State Factorize $\bm\Sigma_{\bm\gamma^{(t)}}$ and compute $\bm v\gets \bm\Sigma_{\bm\gamma^{(t)}}^{-1}\bm r_{\bm\gamma^{(t)}}$
    \State Evaluate $\nabla\mathcal{L}(\bm\gamma^{(t)})$ via~\eqref{eq:grad}
    \State Select $L_t$ by backtracking until~\eqref{eq:bt_accept} holds
    \State $\bm\gamma^{(t+1)}\gets \Pi_{[0,1]^N}\!\big(\bm\gamma^{(t)}-L_t^{-1}\nabla\mathcal{L}(\bm\gamma^{(t)})\big)$
    \State $t\gets t+1$
\Until{the stopping rule is met or $t=T_{\max}$}
\State $\hat{\mathcal{S}}\gets$ indices of the $K$ largest entries of $\bm\gamma^{(t)}$
\end{algorithmic}
\end{algorithm}

\begin{figure*}[!t]
    \centering
    \includegraphics[width=0.7\textwidth]{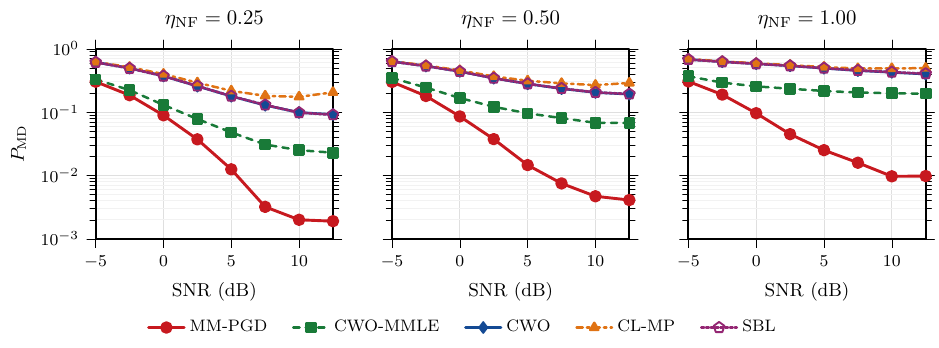}
    \caption{Miss-detection probability $P_{\mathrm{MD}}$ versus SNR for \textbf{MM-PGD} (proposed algorithm) and four baselines under three NF ratios $\eta_{\mathrm{NF}}\in\{0.25,0.50,1.00\}$.
    The operating point is $(N,K,M,L)=(200,30,48,20)$ with $L_n=4$ and LoS-to-scattering power ratio $-5$~dB.}
    \label{fig:exp1}
\end{figure*}

\begin{figure}[!t]
    \centering
    \includegraphics[width=0.7\linewidth]{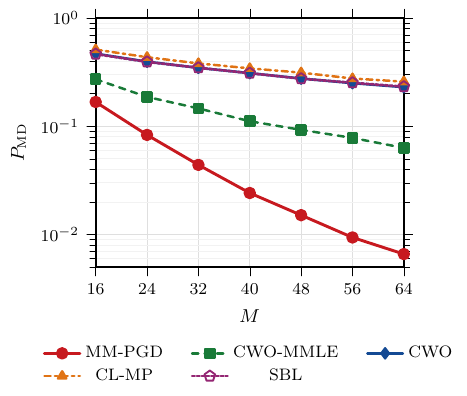}
    \caption{Miss-detection probability $P_{\mathrm{MD}}$ versus antenna count $M$ at $\mathrm{SNR}=5$~dB and fixed $\eta_{\mathrm{NF}}=0.5$.}
    \label{fig:exp2}
\end{figure}

\begin{figure}[!t]
    \centering
    \includegraphics[width=0.7\linewidth]{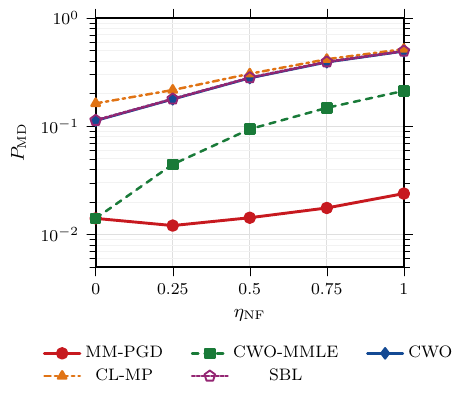}
    \caption{Miss-detection probability $P_{\mathrm{MD}}$ versus NF ratio $\eta_{\mathrm{NF}}$ at $\mathrm{SNR}=5$~dB.}
    \label{fig:exp3}
\end{figure}

% ==========

We employ the Majorization--Minimization (MM) principle to derive a monotonically descending update~\cite{sun2016majorization}.
At iteration~$t$, the Descent Lemma provides the quadratic upper bound
\begin{equation}
\mathcal{L}(\bm\gamma)\le
\underbrace{
\mathcal{L}(\bm\gamma^{(t)})
+
\nabla\mathcal{L}(\bm\gamma^{(t)})^\top(\bm\gamma-\bm\gamma^{(t)})
+
\frac{L_t}{2}\|\bm\gamma-\bm\gamma^{(t)}\|_2^2
}_{\triangleq\; Q(\bm\gamma;\,\bm\gamma^{(t)})},
\label{eq:majorizer}
\end{equation}
valid for any $L_t$ exceeding the Lipschitz constant of $\nabla\mathcal{L}$ on $[0,1]^N$.
Minimizing the surrogate $Q$ over the box constraint yields the closed-form projected gradient step
\begin{equation}
\bm\gamma^{(t+1)}=
\Pi_{[0,1]^N}
\!\left(
\bm\gamma^{(t)}-\frac{1}{L_t}\nabla\mathcal{L}(\bm\gamma^{(t)})
\right).
\label{eq:update}
\end{equation}

Since a tight closed-form expression for $L_{\text{Lip}}$ is generally unavailable, we adopt backtracking: at the first iteration we set $L_t=1$, and at later iterations we warm-start from the previously accepted value; if needed, we double~$L_t$ until the sufficient-decrease condition
\begin{equation}
\mathcal{L}(\bm\gamma^{(t+1)})\le Q(\bm\gamma^{(t+1)};\,\bm\gamma^{(t)})
\label{eq:bt_accept}
\end{equation}
is satisfied.
This guarantees monotone descent of the objective at each iteration.
In our implementation, all objective, gradient, and trace terms are evaluated exactly from a direct Cholesky factorization of $\bm\Sigma_{\bm\gamma}$ at each iterate.
In all experiments, we initialize $\bm\gamma^{(0)}=(K/N)\mathbf{1}_N$, set $T_{\max}=50$, and declare convergence when $\|\bm\gamma^{(t+1)}-\bm\gamma^{(t)}\|_2 < 10^{-12}\max\{\|\bm\gamma^{(t)}\|_2,1\}$ for two consecutive iterations, or when the relative NLL change falls below $10^{-8}$.
For the oracle top-$K$ benchmarking protocol used in the main-text experiments, the binary activity estimate is obtained by selecting the $K$ devices with the largest entries in the final~$\bm\gamma$.
The procedure is summarized in Algorithm~\ref{alg:mmpgd}.

% \paragraph{Per-iteration cost.}
Each MM-PGD iteration requires one Cholesky factorization of $\bm\Sigma_{\bm\gamma}\in\mathbb C^{LM\times LM}$ at cost $\mathcal O((LM)^3)$, plus $\mathcal O(N)$ gradient-component evaluations.
For our evaluated moderate-scale operating point $(L,M)=(20,48)$, this amounts to factorizing a $960\times 960$ matrix, which remains tractable on commodity hardware.
% Backtracking adds at most a small constant number of extra NLL evaluations per iteration (gradient and trace terms are not recomputed).
% When all FF devices, algorithm simplifies:
The same formulation also reveals two explicit special cases. 
If all devices are FF, then every $\bm C_n=g_n(\bm I_M\otimes \bm s_n\bm s_n^\herm)$ and the aggregate covariance reduces to $\bm\Sigma_{\bm\gamma}=\bm I_M\otimes \bm A_{\bm\gamma}$ with $\bm A_{\bm\gamma}=\sum_{n=1}^{N}\gamma_n g_n\bm s_n\bm s_n^\herm+\sigma_w^2\bm I_L$, so exact likelihood evaluation collapses to one $L\times L$ Cholesky factorization and the dominant cost becomes $\mathcal O(L^3)$.
% When all FF devices, algorithm ?:
If all devices are NF, the algorithmic form is unchanged, but the FF simplification is no longer available, so the dominant exact cost stays at $\mathcal O((LM)^3)$.
The mixed NF/FF case treated here is the hybrid generalization of these two extremes, and the all-FF and all-NF scenarios follow directly from the corresponding user covariance matrices.
In the present direct implementation, hybrid NF/FF operation is therefore governed by the general $\mathcal O((LM)^3)$ exact factorization cost, with the all-FF case as the explicit low-complexity reduction and the all-NF case as the no-simplification extreme.

% \begin{figure*}[!t]
%     \centering
%     \includegraphics[width=0.9\textwidth]{figures/fig_exp1.pdf}
%     \caption{Miss-detection probability $P_{\mathrm{MD}}$ versus SNR for \textbf{MM-PGD} (proposed algorithm) and four baselines under three NF ratios $\eta_{\mathrm{NF}}\in\{0.25,0.50,1.00\}$.
%     The operating point is $(N,K,M,L)=(200,30,48,20)$ with $L_n=4$ and LoS-to-scattering power ratio $-5$~dB.}
%     \label{fig:exp1}
% \end{figure*}

% \begin{figure}[!t]
%     \centering
%     \includegraphics[width=0.8\linewidth]{figures/fig_exp2.pdf}
%     \caption{Miss-detection probability $P_{\mathrm{MD}}$ versus antenna count $M$ at $\mathrm{SNR}=5$~dB and fixed $\eta_{\mathrm{NF}}=0.5$.}
%     \label{fig:exp2}
% \end{figure}

% \begin{figure}[!t]
%     \centering
%     \includegraphics[width=0.8\linewidth]{figures/fig_exp3.pdf}
%     \caption{Miss-detection probability $P_{\mathrm{MD}}$ versus NF ratio $\eta_{\mathrm{NF}}$ at $\mathrm{SNR}=5$~dB.}
%     \label{fig:exp3}
% \end{figure}

% =============================================
% =============================================
\section{Numerical Results}
\label{sec:results}

% \begin{figure*}[!t]
%     \centering
%     \includegraphics[width=\textwidth]{figures/fig_exp1.pdf}
%     \caption{$P_{\mathrm{MD}}$ versus SNR for three NF ratios $\eta_{\mathrm{NF}}\in\{0.25,0.50,1.00\}$.
%     The operating point is $(N,K,M,L)=(200,30,48,20)$ with $L_n=4$ and LoS-to-scattering power ratio $-5$~dB.}
%     \label{fig:exp1}
% \end{figure*}

% \begin{figure}[!t]
%     \centering
%     \includegraphics[width=0.9\linewidth]{figures/fig_exp2.pdf}
%     \caption{$P_{\mathrm{MD}}$ versus antenna count $M$ at $\mathrm{SNR}=5$~dB and fixed $\eta_{\mathrm{NF}}=0.5$.}
%     \label{fig:exp2}
% \end{figure}

% \begin{figure}[!t]
%     \centering
%     \includegraphics[width=0.9\linewidth]{figures/fig_exp3.pdf}
%     \caption{$P_{\mathrm{MD}}$ versus near-field composition $\eta_{\mathrm{NF}}$ at $\mathrm{SNR}=5$~dB.}
%     \label{fig:exp3}
% \end{figure}

\subsection{Setup}
We use a mixed NF/FF single-cell setting with carrier frequency $f_c=3$~GHz and cell radius $R=500$~m.
The operating point is
\[
(N,K,M,L)=(200,30,48,20),
\]
with $L_n=4$ scatterers for each NF device, NF ratio $\eta_{\mathrm{NF}}=0.5$, and a fixed LoS-to-scattering power ratio of $-5$~dB, where $\eta_{\mathrm{NF}}$ denotes the fraction of NF devices relative to the full registered device pool.
We generate constant-modulus QPSK pilots and sample the active set uniformly without replacement in each Monte Carlo trial.
Users are drawn uniformly in area from NF and FF annuli determined by the prescribed NF ratio, with i.i.d.\ polar angles in $[0,\pi]$.
Each NF device is assigned $L_n$ scatterers placed uniformly in a disk centered at the BS; their locations define the structured NLoS covariance, while path-loss effects are generated through the distance-based law $\ell(r)=r^{-\alpha_{\mathrm{PL}}}$ with path-loss exponent $\alpha_{\mathrm{PL}}$, and the overall channel power is normalized to satisfy the prescribed LoS-to-scattering ratio.
All results are averaged over $N_{\mathrm{MC}}=500$ Monte Carlo trials.
We compare MM-PGD against four baselines: the coordinatewise optimization algorithm (CWO)~\cite{fengler2021nonTIT}, its mismatched Rician-MLE variant CWO-MMLE~\cite{Liu2024CWOMMLE}, covariance-based matching pursuit (CL-MP)~\cite{marata2024activity}, and sparse Bayesian learning (SBL)~\cite{wipf2004sparse}.
Here CWO-MMLE is the strongest competing covariance-based baseline: it retains the LoS mean $\bar{\bm h}_n$ but replaces the structured covariance by an isotropic surrogate~$g_n\bm I_M$.
We use miss-detection probability $P_{\mathrm{MD}}$ as the main performance metric, and all main-text plots use the oracle top-$K$ operating point for fair support-recovery comparison, as is standard practice for benchmarking the fundamental support-recovery capability of such detectors independently of thresholding heuristics.

\subsection{SNR Sweep}

% \begin{figure*}[!t]
%     \centering
%     \includegraphics[width=\textwidth]{figures/fig_exp1.pdf}
%     \caption{$P_{\mathrm{MD}}$ versus SNR for three NF ratios $\eta_{\mathrm{NF}}\in\{0.25,0.50,1.00\}$.
%     The operating point is $(N,K,M,L)=(200,30,48,20)$ with $L_n=4$ and LoS-to-scattering power ratio $-5$~dB.}
%     \label{fig:exp1}
% \end{figure*}

Figure~\ref{fig:exp1} shows that MM-PGD maintains the lowest $P_{\mathrm{MD}}$ across the full SNR range and all three NF compositions.
At $\eta_{\mathrm{NF}}=0.25$ and $\mathrm{SNR}=10$~dB, MM-PGD achieves $P_{\mathrm{MD}}=0.002$ versus $0.025$ for CWO-MMLE (${\approx}12\times$ gap).
At $\eta_{\mathrm{NF}}=1.00$, CWO-MMLE saturates near $P_{\mathrm{MD}}\approx 0.20$ while MM-PGD continues to improve to $0.010$, which shows a $20\times$ advantage.
This confirms that the gain grows with the proportion of NF devices and is not merely an SNR effect.

\subsection{Antenna-Count Sweep}

% \begin{figure}[!t]
%     \centering
%     \includegraphics[width=\linewidth]{figures/fig_exp2.pdf}
%     \caption{$P_{\mathrm{MD}}$ versus antenna count $M$ at $\mathrm{SNR}=5$~dB and fixed $\eta_{\mathrm{NF}}=0.5$.}
%     \label{fig:exp2}
% \end{figure}

Figure~\ref{fig:exp2} fixes the NF/FF composition at $\eta_{\mathrm{NF}}=0.5$ and varies the array size.
MM-PGD improves from $P_{\mathrm{MD}}=0.168$ ($M=16$) to $0.007$ ($M=64$), a $24\times$ reduction.
CWO-MMLE tracks a similar trend but remains consistently higher, decreasing from $0.272$ to $0.063$ as $M$ grows.
The remaining baselines stay in the range $P_{\mathrm{MD}}\approx 0.23$--$0.51$ and improve only modestly, confirming that structured covariance exploitation becomes increasingly valuable as the array grows.

\subsection{NF-Ratio Sweep}

% \begin{figure}[!t]
%     \centering
%     \includegraphics[width=\linewidth]{figures/fig_exp3.pdf}
%     \caption{$P_{\mathrm{MD}}$ versus near-field composition $\eta_{\mathrm{NF}}$ at $\mathrm{SNR}=5$~dB.}
%     \label{fig:exp3}
% \end{figure}

Figure~\ref{fig:exp3} isolates the impact of the NF fraction.
At $\eta_{\mathrm{NF}}=0$ (all FF), MM-PGD and CWO-MMLE both achieve $P_{\mathrm{MD}}=0.014$, confirming that their advantage is driven by NF structure rather than algorithmic overhead.
As $\eta_{\mathrm{NF}}$ increases, CWO-MMLE degrades to $0.213$ at $\eta_{\mathrm{NF}}=1$ ($15\times$ increase), while MM-PGD remains nearly flat at $0.024$ ($1.7\times$).
The takeaway is once structured NF covariance dominates, full-vector covariance-aware optimization is substantially more effective than coordinate-wise or isotropic alternatives.

\section{Conclusion}
\label{sec:conclusion}

Mixed NF/FF activity detection under a unified Rician model and proposed a covariance-aware MM-PGD detector with full-vector updates has been addressed in the paper. 
Numerical results showed that MM-PGD performs on par with CWO-MMLE in the all-FF limit and achieves clear gains as the NF ratio grows, which indicates that its benefit comes from exploiting structured NF covariance rather than from a generic optimization effect. 
Developing more scalable implementations of the same covariance-aware formulation for larger arrays and larger device pools is a future work. 
A particular focus will be on reducing the cost of exact likelihood evaluation in strongly coupled hybrid NF/FF settings.

% From a detector-design perspective, the main message is that mixed NF/FF propagation should not be treated as a minor perturbation of the classical FF model. When a noticeable fraction of devices lies in the near field, isotropic covariance surrogates discard useful structure, and coordinate-wise updates become less attractive because each NF device induces a higher-rank covariance perturbation. In contrast, the proposed MM-PGD update operates directly on the globally coupled relaxed likelihood and converts the additional NF structure into a detection gain.
%
% The present study assumes known long-term statistics and a fixed NF/FF partition, with an offline/online split where the per-device mean vectors and covariance matrices are precomputed for the registered user pool. Future work includes lower-complexity scalable implementations, robust or online learning of the long-term statistical CSI $\{(\bar{\bm h}_n,\bm R_n)\}_{n=1}^N$, sensing-assisted NF/FF inference in integrated sensing and communications (ISAC), threshold-free detection without an oracle $K$, and extension to wideband or multi-cell scenarios.

\bibliographystyle{IEEEtran}
\bibliography{IEEEabrv,reference}

\end{document}